\documentclass[11pt]{article}

\usepackage[english]{babel}								
\usepackage[utf8]{inputenc}								
\usepackage{mathptmx}
\usepackage[T1]{fontenc}								 
\usepackage{amsmath,amsfonts,amssymb,amsthm,cancel,siunitx,
	calculator,calc,mathtools,empheq,latexsym}

\usepackage{subfig,epsfig,tikz,float}		            
\usepackage{booktabs,multicol,multirow,tabularx,array}    
\usepackage[hidelinks]{hyperref}
\hypersetup{colorlinks=true,linkcolor=blue, citecolor=blue, urlcolor=blue}
\usepackage{titlesec}

\titleformat*{\section}{\Large\bfseries}
\title{\Large\bf
	{The S-matrix Bootstrap: From the Sine-Gordon Model to Celestial Amplitudes}}

\author{Valeriia Stolbova }	

\begin{document}
\numberwithin{equation}{section}
\thispagestyle{empty}
	\date{}
	\maketitle
	\vspace{-1cm}
	\begin{center}
		{\normalsize
			Steklov Mathematical Institute, Fontanka 27, St. Petersburg, 191023, Russia \\
			\vspace{5mm}
			E-mail: stolbova@pdmi.ras.ru}
	\end{center}

	\bigskip
	\noindent
	{\small \centering{{\bf Abstract}}\\
		We address the scattering problem in two-dimensional integrable models, focusing on the sine-Gordon theory. We use the S-matrix bootstrap approach based on analytical properties of the S-matrix to compute scattering amplitudes of the sine-Gordon model. We obtain original integral representations for the sine-Gordon scattering amplitudes involving bound states. Calculation of minimal form factors in this model follows directly from integral representations of scattering amplitudes. We also propose the extension of the sine-Gordon bootstrap equations to the space of conformal correlators known as celestial amplitudes.}
	
	
\newpage
\small\tableofcontents
\normalsize

\setlength{\jot}{2pt}
\section{Introduction}

The S-matrix program is an analytic non-perturbative approach to the calculation of scattering amplitudes in two-dimensional integrable models \cite{chew,coleman}.
The main purpose of this program is to formulate a self-consistent mathematical framework that would classify integrable models according to their analytic properties and provide their exact solutions in terms of the S-matrix operator, dating back to the original works \cite{heis, eden}. In this approach, the importance of studying analytic continuations of amplitudes is underlined, and the attempts to calculate scatterring amplitudes without the use of a Lagrangian are presented.

This program was mainly developed in the context of two-dimensional quantum field theories, where the S-matrices are non-trivial and can be obtained explicitly due to factorization \cite{parke,iag}. The factorization property of integrable S-matrices, including that of the sine-Gordon model \cite{aref_kor}, together with its analytical structure provide enough restrictions to obtain exact expressions for the scattering amplitudes \cite{kttw}. 
The essential algebraic structure describing the S-matrix properties is the so-called Zamolodchikov-Faddeev algebra \cite{zamzam}. In this approach, fundamental particles and bound states are treated on the same footing, giving rise to the corresponding additional relations in the algebra. These relations are analogous to the conformal bootstrap operator expansions \cite{bpz}. 

In the context of the bootstrap program for integrable quantum field theories, the matrix elements of local operators known as form factors have been extensively studied \cite{branch,sff}. Further applications include computation of correlation functions via infinite summation over the multi-particle form factors. Form factors of the sine-Gordon model are commonly given in terms of integral representations that can be obtained directly from certain integral representations of corresponding scattering amplitudes \cite{ssgff}. The form factor program is then developed using, for instance, off-shell Bethe ansatz technique \cite{b_k_sg_ff}. Investigations of the sine-Gordon bound state form factors are closely connected to the study of the sinh-Gordon form factors, since at the Lagrangian level the two models are related by the complexification of the coupling constant \cite{ffsinh, branch}. 

Recent research in the context of the correspondence between scattering amplitudes and correlators in conformal field theory proposed further extension of the bootstrap principle. Scattering amplitudes written in a so-called conformal basis are interpreted as conformal correlators known as celestial amplitudes \cite{conf_bas}. This interpretation is motivated by the study of conformal symmetries arising in higher-dimensional scattering amplitudes that was recently verified in the context of four-dimensional quantum gravity \cite{quant_gr}. Recent research \cite{cel_2d, kap_trop} demonstrated that celestial amplitudes corresponding to the two-dimensional integrable S-matrices obtain a particular simplified form and, therefore, it is possible to use the bootstrap principle to study conformal correlators in this approach.

In this paper we study the bootstrap approach to the sine-Gordon scattering data. We provide the analytical constraints sufficient to compute  scattering amplitudes of the main sine-Gordon spectrum. These constraints include the factorization property of the S-matrix, analytical properties of scattering amplitudes and the Zamolodchikov-Faddeev algebraic relations. We use the bootstrap method analogous to conformal bootstrap to compute integral representations of the sine-Gordon scattering amplitudes involving bound states. Consistency checks show that our results correspond to the known trigonometric expressions for these amplitudes. We also investigate the applications of bootstrap approach to certain correlation functions. We review the connection between the sine-Gordon scattering amplitudes in their integral form and minimal form factors. We also address the bootstrap problem in the context of conformal correlators known as celestial amplitudes. As for celestial amplitudes corresponding to two-dimensional integrable models, a full set of bootstrap constraints has not been established yet. We provide the constraints for celestial amplitudes corresponding to the sine-Gordon analytical properties and the bootstrap equation.

This paper is organized as follows. In Section 2 we review the general analytical properties of the two-dimensional integrable S-matrices and provide the S-matrix of the sine-Gordon model obtained from the Zamolodchikov-Faddeev algebraic formalism. In Section 3 we use the bootstrap method to compute original integral representations of the sine-Gordon amplitudes involving bound states. In Section 4 we discuss the connection between scattering amplitudes and correlation functions. We obtain minimal form factors from integral scattering amplitudes and translate analytical constraints corresponding to the sine-Gordon bootstrap to the space of celestial amplitudes.

\bigskip

\section{S-matrix of the sine-Gordon model from analytical constraints}

In this section we approach the sine-Gordon scattering problem within the S-matrix program framework. We provide the set of analytical constraints including factorization present in two-dimensional integrable models, general properties of the S-matrix and Zamolodchikov-Faddeev algebraic formalism. These constraints are sufficient to determine the scattering amplitudes for the main sine-Gordon spectrum. Another constraint required for theories with bound states, such as the sine-Gordon model, is the bootstrap equation originating from conformal bootstrap. The bootstrap approach to the total sine-Gordon scattering matrix, its applications and generalizations constitute the main interest of this study.

\subsection{Factorized scattering}

The S-matrices in two-dimensional integrable models, where high-order conserved charges are present, are not only non-trivial, but also exactly solvable \cite{kttw}. The $n$-particle S-matrix corresponding to elastic scattering process is generally defined as a unitary linear operator
\begin{equation}
	S_n\left(p_1,\dots,\;p_n\right)={}_{out} {\left\langle p_1,\dots,p_n \;\vert\; p_1,\dots,p_n \right\rangle}_{in}
\end{equation}
The general $n$-particle S-matrix of a two-dimensional integrable model posesses the crucial property of factorizing into the product of $n\left(n-1\right)/2$ two-particle ones \cite{iag}
\begin{equation}
	{{S}_{n}}\left( {{p}_{1}},\ldots,{{p}_{n}} \right)=\prod\limits_{i=1}^{n-1}{\prod\limits_{j=i+1}^{n}{{{S}_{2}}\left( {{p}_{i}},{{p}_{j}} \right)}}.
\end{equation}
In two dimensions, it is convenient to use the parametrization in terms of the rapidity variable $\theta$:
\begin{equation}
	p_{i}^{0}\,=\,{{m}_{i}}\,\cosh \,{{\theta }_{i}},\qquad p_{i}^{1}\,=\,{{m}_{i}}\,\sinh \,{{\theta }_{i}}.
\end{equation}
Due to Lorentz invariance, we consider scattering amplitudes to be functions of rapidity differences ${{\theta }_{ij}}={{\theta }_{i}}-{{\theta }_{j}}\,.$
For the case of factorized scattering, the function $S \left( \theta \right)$ is meromorphic in the complex $\theta$-plane  (here $\theta \equiv {\theta}_{12}$) and obtains real values only on the imaginary $\theta$-axis. The region $0<\operatorname{Im}\theta<\pi$ corresponds to the physical sheet in terms of rapidity. If bound states are present in the model, they correspond to simple poles of the $S$-matrix in the given range of $\theta$ \cite{zamzam}.

\subsection{Zamolodchikov-Faddeev algebra}

Let us describe an algebraic setup that is consistent with the analytical properties of the S-matrix. 

We introduce the following expression for asymptotic states using a set of vertex operators  ${{A}_{{{a}_{i}}}}\left( {{p}_{i}} \right),\;\; {{A}^{\dagger }}^{{{a}_{i}}}\left( {{p}_{i}} \right)$ that generalize creation and annihilation operators
\begin{equation}
	\label{algasymp}
	\begin{aligned}
		& \left| {{\theta }_{1}},{{\theta }_{2}},\ldots {{\theta }_{n}} \right\rangle _{{{a}_{1}},\,{{a}_{2}},\ldots ,{{a}_{n}}}^{in}\,={{A}_{{{a}_{1}}}}\left( {{\theta }_{1}} \right){{A}_{{{a}_{2}}}}\left( {{\theta }_{2}} \right)\ldots {{A}_{{{a}_{n}}}}\left( {{\theta }_{n}} \right)\left| 0 \right\rangle ,\\ 
		& \left| {{\theta }_{1}},{{\theta }_{2}},\ldots {{\theta }_{n}} \right\rangle _{{{a}_{1}},{{a}_{2}},\ldots ,{{a}_{n}}}^{out}={{A}_{{{a}_{n}}}}\left( {{\theta }_{n}} \right)\ldots {{A}_{{{a}_{2}}}}\left( {{\theta }_{2}} \right){{A}_{{{a}_{1}}}}\left( {{\theta }_{1}} \right)\left| 0 \right\rangle, \\ 
		& {}_{{{a}_{1}},{{a}_{2}},\ldots,{{a}_{n}}}^{in}\left\langle  {{\theta }_{1}},{{\theta }_{2}},\ldots {{\theta }_{n}} \right|=\left\langle  0 \right|{{A}^{\dagger }}^{{{a}_{1}}}\left( {{\theta }_{1}} \right)\,{{A}^{\dagger }}^{{{a}_{2}}}\left( {{\theta }_{2}} \right)\ldots {{A}^{\dagger }}^{{{a}_{n}}}\left( {{\theta }_{n}} \right),\\ 
		& {}_{{{a}_{1}},{{a}_{2}},\ldots ,{{a}_{n}}}^{out}\left\langle  {{\theta }_{1}},{{\theta }_{2}},\ldots {{\theta }_{n}} \right|=\left\langle  0 \right|{{A}^{\dagger }}^{{{a}_{n}}}\left( {{\theta }_{n}} \right)\ldots{{A}^{\dagger }}^{{{a}_{2}}}\left( {{\theta }_{2}} \right){{A}^{\dagger }}^{{{a}_{1}}}\left( {{\theta }_{1}} \right). 
	\end{aligned}
\end{equation}
Here the indices $a_i$ label the particle type and we specify the decreasing order of rapidities as $\theta_{1} > \theta_{2} > \theta_{3} > \dots > \theta_n$.

The operators ${{A}_{{{a}_{i}}}}\left( {{p}_{i}} \right),\;\; {{A}^{\dagger }}^{{{a}_{i}}}\left( {{p}_{i}} \right)$ form an associative non-commutative Zamolodchikov-Faddeev algebra \cite{zamzam}.  The definition of asymptotic states given above makes it possible to interpret scattering processes as permutations of the corresponding vertex operators. The defining commutation relations of the Zamolodchikov-Faddeev algebra are of the form
\begin{equation}
	\label{algcommrel}
	\begin{aligned}
		{{A}_{i}}\left( {{\theta }_{1}} \right){{A}_{j}}\left( {{\theta }_{2}} \right) &={{A}_{l}}\left( {{\theta }_{2}} \right){{A}_{k}}\left( {{\theta }_{1}} \right)S_{ij}^{kl}\left( {{\theta }_{1}}-{{\theta }_{2}} \right), \\ 
		{{A}^{\dagger i}}\left( {{\theta }_{1}} \right){{A}^{\dagger j }}\left( {{\theta }_{2}} \right) &=S_{kl}^{ij}\left( {{\theta }_{1}}-{{\theta }_{2}} \right){{A}^{\dagger l}}\left( {{\theta }_{2}} \right){{A}^{\dagger k}}\left( {{\theta }_{1}} \right), \\ 
		{{A}^{\dagger k}}\left( {{\theta }_{1}} \right){{A}_{j}}\left( {{\theta }_{2}} \right) &={{A}_{l}}\left( {{\theta }_{2}} \right)S_{ij}^{kl}\left( {{\theta }_{2}}-{{\theta }_{1}} \right){{A}^{\dagger i}}\left( {{\theta }_{1}} \right)+\delta \left( {{\theta }_{1}}-{{\theta }_{2}} \right)\delta _{j}^{k}. 
	\end{aligned}
\end{equation}
These relations generalize the usual bosonic and fermionic commutation relations.

\subsection{S-matrix bootstrap}

The S-matrix bootstrap method uses the assumption that poles of the two-particle amplitude in the physical region correspond to the bound states of the model \cite{b_k_sg_ff}.

Given that a bound state $n$ is formed by particles $i$ and $j$, it is interpreted as the two-particle amplitude pole at the value of rapidity given by
\begin{equation}
	\theta=iu_{ij}^{n},
\end{equation}
and the corresponding amplitude can be wriiten in a form \cite{muss,bomb}
\begin{equation}
	S_{ij}^{kl}\left( \theta  \right)\simeq \frac{\Gamma _{ij}^{n}\,{{R}_{n}}\,\Gamma _{n}^{kl}}{\theta -iu_{ij}^{n}},
\end{equation}
where ${R}_{n}$ is the residue and $\Gamma _{ij}^{n},\; \Gamma _{n}^{kl}$ are projection operators onto the space of the bound state $n$.

For the case of diagonal S-matrices that we consider further, the bootstrap equation expresses the S-matrix for any particle $k$ and a bound state $n$ simply as 
\begin{equation}
	\label{bootdiag0}
	{{S}_{kn}}\left( \theta  \right)={{S}_{ki}}\left( \theta -i\bar{u}_{in}^{j} \right){{S}_{kj}}\left( \theta +i\bar{u}_{jn}^{i} \right),
\end{equation}
where $\bar{u}=\pi-u$ is the angle introduced for further convenience.

\bigskip

\subsection{General O(N)-symmetric S-matrix}

The spectrum of an integrable model possessing $O(N)$ isotopic symmetry is given by $n$ partices ${{A}_{i}},\quad i=1,2,\ldots N$ of equal masses $m$ \cite{zamzam}.

\setlength{\jot}{4pt}
The general S-matrix takes the form
\begin{equation}
	\begin{aligned}
		\label{s_on}
		_{ik}{{S}_{jl}}&=\left\langle {{A}_{j}}\left( {{p}_{1}}^{\prime } \right){{A}_{l}}\left( {{p}_{2}}^{\prime } \right),out|{{A}_{i}}\left( {{p}_{1}} \right){{A}_{k}}\left( {{p}_{2}} \right),in \right\rangle\\ 
		&=\delta \left( {{p}_{1}}-{{p}_{1}}^{\prime } \right)\delta \left( {{p}_{2}}-{{p}_{2}}^{\prime } \right)\left[ {{\delta }_{ik}}{{\delta }_{jl}}{{S}_{1}}\left( s \right)+{{\delta }_{ij}}{{\delta }_{kl}}{{S}_{2}}\left( s \right)+{{\delta }_{il}}{{\delta }_{jk}}{{S}_{3}}\left( s \right) \right] \\
		&\hspace{10pt}\pm \delta \left( {{p}_{2}}-{{p}_{1}}^{\prime } \right)\delta \left( {{p}_{1}}-{{p}_{2}}^{\prime } \right)\left[ {{\delta }_{ik}}{{\delta }_{jl}}{{S}_{1}}\left( s \right)+{{\delta }_{kj}}{{\delta }_{il}}{{S}_{2}}\left( s \right)+{{\delta }_{kl}}{{\delta }_{ij}}{{S}_{3}}\left( s \right) \right],
	\end{aligned}
\end{equation}
\setlength{\jot}{2pt}
where $s$ is the usual Mandelstam variable $s=\left(p_1+p_2\right)^2$ and $s=4{{m}^{2}}\,{{\cosh }^{2}}\left( {\theta }/{2}\right)$.

Now we outline the main properties of the S-matrix in terms of rapidity.

Crossing relations take the form
\begin{equation}
	\label{cross}
	\begin{aligned}
		& {{S}_{2}}\left( \theta  \right)={{S}_{2}}\left( i\pi -\theta  \right),\\ 
		& {{S}_{1}}\left( \theta  \right)={{S}_{3}}\left( i\pi -\theta  \right).
	\end{aligned}
\end{equation}
Unitarity is expressed by
\begin{multline}
	\label{unit}
	{{S}_{2}}\left( \theta  \right){{S}_{2}}\left( -\theta  \right)+{{S}_{3}}\left( \theta  \right){{S}_{3}}\left( -\theta  \right)=1,\\ 
	\shoveleft{{{S}_{2}}\left( \theta  \right){{S}_{3}}\left( -\theta  \right)+{{S}_{2}}\left( -\theta  \right){{S}_{3}}\left( \theta  \right)=1},\\ 
	\shoveleft{N{{S}_{1}}\left( \theta  \right){{S}_{1}}\left( -\theta  \right)+{{S}_{1}}\left( \theta  \right){{S}_{2}}\left( -\theta  \right)+{{S}_{1}}\left( \theta  \right){{S}_{3}}\left( -\theta  \right)}\\
	+{{S}_{2}}\left( \theta  \right){{S}_{1}}\left( -\theta  \right)+{{S}_{3}}\left( \theta  \right){{S}_{1}}\left( -\theta  \right)=0.
\end{multline}
Commutation relations for the operators ${{A}_{i}}\left( \theta  \right),\quad i=1,2,\ldots N$ corresponding to the given symmetry can be written as
\begin{multline}
	\label{comm}
	{{A}_{i}}\left( {{\theta }_{1}} \right){{A}_{j}}\left( {{\theta }_{2}} \right)={{\delta }_{ij}}{{S}_{1}}\left( {{\theta }_{12}} \right)\sum\nolimits_{k=1}^{N}{{{A}_{k}}\left( {{\theta }_{2}} \right){{A}_{k}}}\left( {{\theta }_{1}} \right),\\
	+{{S}_{2}}\left( {{\theta }_{12}} \right){{A}_{j}}\left( {{\theta }_{2}} \right){{A}_{i}}\left( {{\theta }_{1}} \right)
	+{{S}_{3}}\left( {{\theta }_{12}} \right){{A}_{i}}\left( {{\theta }_{2}} \right){{A}_{j}}\left( {{\theta }_{1}} \right).
\end{multline}
Since the sine Gordon model possesses $O(2)$-symmetry, we restrict our further discussion to this case only, since the solutions for models with $O(N)$-symmetry with $N\ge3$ differ significantly from the $O(2)$ case.

Factorization equations in this case are written as
\begin{multline}
	\label{fact2}
	{{S}_{2}}{{S}_{1}}{{S}_{3}}+{{S}_{2}}{{S}_{3}}{{S}_{3}}+{{S}_{3}}{{S}_{3}}{{S}_{2}}={{S}_{3}}{{S}_{2}}{{S}_{3}}+{{S}_{1}}{{S}_{2}}{{S}_{3}}+{{S}_{1}}{{S}_{1}}{{S}_{2}},\\
	\shoveleft{{{S}_{3}}{{S}_{1}}{{S}_{3}}+{{S}_{3}}{{S}_{2}}{{S}_{3}}={{S}_{3}}{{S}_{3}}{{S}_{1}}+{{S}_{3}}{{S}_{3}}{{S}_{2}}+{{S}_{2}}{{S}_{3}}{{S}_{1}}+{{S}_{2}}{{S}_{3}}{{S}_{3}}}\\
	+2{{S}_{1}}{{S}_{3}}{{S}_{1}}+{{S}_{1}}{{S}_{3}}{{S}_{2}}+{{S}_{1}}{{S}_{3}}{{S}_{3}}+{{S}_{1}}{{S}_{2}}{{S}_{1}}+{{S}_{1}}{{S}_{1}}{{S}_{1}}. 
\end{multline}
Here the arguments are given by $\theta ,\; \theta +{\theta }',\; {\theta }',$ respectively in each term.

Solutions of these equations, together with the crossing symmetry and unitarity conditions for the $O(2)$ case are given by
\begin{equation}
	\label{sol2}
	\begin{aligned}
		& {{S}_{3}}\left( \theta  \right)=i\cot \left( \frac{4\pi^2 }{\gamma } \right)\coth \left( \frac{4\pi \theta }{\gamma } \right){{S}_{2}}\left( \theta  \right) \\ 
		& {{S}_{1}}\left( \theta  \right)=i\cot \left( \frac{4\pi^2 }{\gamma } \right)\coth \left( \frac{4\pi \left( i\pi -\theta  \right)}{\gamma } \right){{S}_{2}}\left( \theta  \right) \\ 
	\end{aligned}
\end{equation}
and
\begin{equation}
	\label{fromu2}
	\begin{aligned}
		&{{S}_{2}}\left( \theta  \right){{S}_{2}}\left( -\theta  \right)=\frac{{{\sin }^{2}}\left( \frac{4\pi^2 }{\gamma } \right){{\sinh }^{2}}\left( \frac{4\pi \theta }{\gamma } \right)}{{{\sin }^{2}}\left( \frac{4\pi^2 }{\gamma } \right){{\sinh }^{2}}\left( \frac{4\pi \theta }{\gamma } \right)+{{\cos }^{2}}\left( \frac{4\pi^2}{\gamma } \right){{\cosh }^{2}}\left( \frac{4\pi \theta }{\gamma } \right)}.\\
	\end{aligned}
\end{equation}

General form of $S_2\left(\theta\right)$ up to a CDD ambiguity is specified in the work \cite{zamzam}. 
It has been established that the solution ${{S }_{2}} \left( \theta \right)$ corresponds to the exact S-matrix of the sine-Gordon model with the Lagrangian
\begin{equation}
	\label{SG}
	{{\mathcal{L}}_{SG}}=\frac{1}{2}{{\left( {{\partial }_{\mu }}\,\phi  \right)}^{2}}+\frac{{{m}_{0}}^{2}}{{{\beta }^{2}}}\cos \left( \beta \phi  \right).
\end{equation}
It is important to note that the sine-Gordon model coupling constant $\beta$ is the analytic continuation of the sinh-Gordon model coupling constant, and the scattering matrices of these models are also related \cite{intsys}. Factorization property of the sine-Gordon S-matrix has been demonstrated by direct perturbative calculations in \cite{aref_kor}.

\subsection{Algebraic relations for the sine-Gordon S-matrix}	 

Considering the sine-Gordon model, it follows from $O\left(2\right)$ symmetry that a soliton and an antisoliton form an isovector $O\left(2\right)$-doublet with real components ${A}_{i},\quad\,i=1,2,$ and consequently can be expressed as \cite{zamzam}, \cite{zamsg}
\begin{equation}
	A\left( \theta \right) = {A}_{1} \left( \theta \right) + i{A}_{2}\left( \theta \right),\qquad \bar{A}\left( \theta \right) = {A}_{1} \left( \theta \right) - i{A}_{2}\left( \theta \right).
\end{equation}
Commutation relations for particles  $A\left( \theta \right)$ and $\bar{A} \left( \theta \right)$ are constrained to be
\begin{equation}
	\begin{aligned}
		A\left( {\theta}_{1} \right) \bar{A}\left( {\theta}_{2} \right) &= {S}_{T} \left( {\theta}_{12} \right) \bar{A} \left( {\theta}_{2} \right) A\left( {\theta}_{1} \right) + {S}_{R} \left( {\theta}_{12} \right) A\left( {\theta}_{2} \right) \bar{A} \left( {\theta}_{1} \right),\\
		A\left( {\theta}_{1} \right) A\left( {\theta}_{2} \right) &= S \left( {\theta}_{12} \right) A \left( {\theta}_{2} \right) A\left( {\theta}_{1} \right), \\
		\bar {A}\left( {\theta}_{1} \right) \bar{A} \left( {\theta}_{2} \right) &= S \left( {\theta}_{12} \right) \bar{A} \left( {\theta}_{2} \right) \bar{A} \left( {\theta}_{1} \right), \\
	\end{aligned}
\end{equation}
For the soliton-soliton amplitude 
$S\left( \theta \right)$, the soliton-antisoliton transition amplitude ${S}_{T} \left( \theta \right)$ and the soliton-antisoliton reflection amplitude ${S}_{R} \left( \theta \right)$ are given by
\begin{equation}
	\begin{aligned}
		S\left( \theta \right) &= {S}_{3}\left( \theta \right)+ {S}_{2}\left( \theta \right),\\
		{S}_{T}\left( \theta \right) &= {S}_{1}\left( \theta \right)+ {S}_{2}\left( \theta \right),\\
		{S}_{R}\left( \theta \right) &= {S}_{1}\left( \theta \right)+ {S}_{3}\left( \theta \right).\\
	\end{aligned}
\end{equation}
Unitarity condition can be written as
\begin{equation}
	\label{unit_sg}
	\begin{aligned}
		S\left( \theta  \right)S\left( -\theta  \right) &=1, \\ 
		{{S}_{T}}\left( \theta  \right){{S}_{T}}\left( -\theta  \right)+{{S}_{R}}\left( \theta  \right){{S}_{R}}\left( -\theta  \right) &=1, \\ 
		{{S}_{T}}\left( \theta  \right){{S}_{R}}\left( -\theta  \right)+{{S}_{R}}\left( \theta  \right){{S}_{T}}\left( -\theta  \right) &=0, \\ 
	\end{aligned}
\end{equation}
and crossing symmetry takes the form
\begin{equation}
	\label{cross_sg}
	\begin{aligned}
		S\left( \theta  \right) &={{S}_{T}}\left( i\pi -\theta  \right),\\ 
		{{S}_{R}}\left( \theta  \right) &={{S}_{R}}\left( i\pi -\theta  \right). \\ 
	\end{aligned}
\end{equation}
We introduce the notation
\begin{equation}
	\label{const_sg}
	\begin{aligned}
		& \nu \equiv \frac{8\pi }{{{\beta }^{2}}}-1=\frac{8\pi }{\gamma }.\\
	\end{aligned}
\end{equation}
Due to its factorization and symmetry properties, the sine-Gordon S-matrix corresponds to the following trigonometric solution of the Yang-Baxter equation \cite{zamzam}
\begin{equation}
	\label{sol_sg1}
	\begin{aligned}
		{{S}_{T}}\left( \theta  \right) &=\frac{\sinh \left( \nu \theta  \right)}{\sinh \left( \nu \left( i\pi -\theta  \right) \right)}S\left( \theta  \right), \\ 
		{{S}_{R}}\left( \theta  \right) &=\frac{i\sin \left( \pi \nu  \right)}{\sinh \left( \nu \left( i\pi -\theta  \right) \right)}S\left( \theta  \right). \\ 
	\end{aligned}
\end{equation}
It can also be deduced from crossing symmetry and unitarity relation that $S\left(\theta\right)$ admits the following representation
\begin{multline}
	\label{s_sg_g}
	S\left( \theta \right)=-\prod\limits_{k=0}^{\infty }{\frac{\Gamma \left( 1+\left( 2k+1 \right)\nu -i\frac{\nu \theta }{\pi } \right)\Gamma \left( 1+2k\nu +i\frac{\nu \theta }{\pi } \right)}{\Gamma \left( 1+\left( 2k+1 \right)\nu +i\frac{\nu \theta }{\pi } \right)\Gamma \left( 1+2k\nu -i\frac{\nu \theta }{\pi } \right)}} \\
	\times \frac{\Gamma \left( \left( 2k+1 \right)\nu -i\frac{\nu \theta }{\pi } \right)\Gamma \left( \left( 2k+2 \right)\nu +i\frac{\nu \theta }{\pi } \right)}{\Gamma \left( \left( 2k+1 \right)\nu +i\frac{\nu \theta }{\pi } \right)\Gamma \left( \left( 2k+2 \right)\nu -i\frac{\nu \theta }{\pi } \right)}, 
\end{multline}
where the overall minus sign results from the sine-Gordon scattering properties \cite{bomb}.

\section{Bootstrapping sine-Gordon S-matrix with bound states}

In this section we use the bootstrap method to compute scattering amplitudes for the total sine-Gordon S-matrix. We obtain original integral representations for the amplitudes corresponding to arbitrary soliton-bound state and bound state-bound state scattering processes. We provide some consistency checks and demonstrate the equivalence of obtained results and the existing trigonometric form of the amplitudes. 

\subsection{Total sine-Gordon S-matrix}

Total sine-Gordon S-matrix involves elements that describe scattering between solitons and bound states. The latter paricles appear when $\beta^{2}<4\pi$, and we denote the corresponding algebra elements by ${{B}_{n}}$, where $n=1,2,\dots,\nu.$ 

Then the total S-matrix is of the form
\begin{equation}
	\label{s_full_sg}
	{S}_{SG} \left( \theta  \right)=\left( \begin{matrix}
		S\left( \theta  \right) & {} & {} & {} & {} & {} & {}  \\
		{} & {{S}_{T}}\left( \theta  \right) & {{S}_{R}}\left( \theta  \right) & {} & {} & {} & {}  \\
		{} & {{S}_{R}}\left( \theta  \right) & {{S}_{T}}\left( \theta  \right) & {} & {} & {} & {}  \\
		{} & {} & {} & S\left( \theta  \right) & {} & {} & {}  \\
		{} & {} & {} & {} & {{S}_{s{{b}_{1}}}} & {} & {}  \\
		{} & {} & {} & {} & {} & {{S}_{{{b}_{1}}{{b}_{1}}}} & {}  \\
		{} & {} & {} & {} & {} & {} & \ddots   \\
	\end{matrix} \right). 
\end{equation}
Bound state masses are given by \cite{zamsg}
\begin{equation}
	\label{boundmasssg}
	{m}_{n}=2m \sin \frac {n}{2\nu},\quad\,n=1,2,\,\dots\,,\nu,
\end{equation}
where $m$ is the soliton mass.

Imposing the constraints listed above is sufficient to obtain trigonometric expressions for the amplitudes \cite{zamzam, zamsg}. The results are
\begin{equation}
	\label{zamsbn}
	{{S}_{s{{b}_{n}}}}=\frac{\sinh \theta +i\cos \frac{n\gamma}{16}}{\sinh \theta -i\cos \frac{n\gamma}{16}}\,\,\,\prod\limits_{i=1}^{n-1}{\,\,\frac{{{\sin}^{2}}\left( \frac{n-2l}{32}\gamma -\frac{\pi }{4}+i\frac{\theta}{2} \right)}{{{\sin }^{2}}\left( \frac{n-2l}{32}\gamma -\frac{\pi }{4}-i\frac{\theta }{2} \right)}}
\end{equation}
for the soliton-bound state scattering amplitude ${S}_{s{{b}_{n}}}$, and
\begin{multline}
	\label{zambmbn}
	{{S}_{{{b}_{m}}{{b}_{n}}}}=\frac{\sinh \theta +i\sin \left( \frac{n+m}{16} \gamma\right)}{\sinh \theta -i\sin \left( \frac{n+m}{16}\gamma \right)}\frac{\sinh \theta +i\sin \left( \frac{n-m}{16}\gamma \right)}{\sinh \theta -i\sin \left( \frac{n-m}{16}\gamma \right)} \\ 
	\times \prod\limits_{l=1}^{m-1}{\,\,\frac{{{\sin }^{2}}\left( \frac{m-n-2l}{32}\gamma +i\frac{\theta }{2} \right)}{{{\sin }^{2}}\left( \frac{m-n-2l}{32}\gamma -i\frac{\theta }{2} \right)}}\,\,\frac{{{\cos }^{2}}\left( \frac{m-n-2l}{32}\gamma +i\frac{\theta }{2} \right)}{{{\cos }^{2}}\left( \frac{m-n-2l}{32}\gamma -i\frac{\theta }{2} \right)},\quad n\ge m 
\end{multline}
for the bound state-bound state scattering amplitude $\left(B_m,\; B_n\right)$.

In our paper, we use the bootstrap method to obtain integral representations of such amplitudes and provide some consistency checks with the trigonometric representations given above.

\subsection{Application of the S-matrix bootstrap method}

In the diagonal scattering case, the S-matrix can be computed via bootstrap according to the formula (\ref{bootdiag0}). For the sine-Gordon model where $i{u}_{ij}^{n} = \alpha_{n};\quad i\bar{u}_{in}^{j} = i\bar{u}_{jn}^{i} = {{\alpha}_{n}}/{2}$ we have 
\begin{equation}
	\label{bootdiag}
	\begin{aligned}
		& {{S}_{kn}}\left( \theta  \right)={{S}_{ki}}\left( \theta -\frac{{\alpha}_{n} }{2} \right){{S}_{kj}}\left( \theta +\frac{{\alpha}_{n} }{2} \right), \\ 
	\end{aligned}
\end{equation}
where $n$ is the bound state formed by particles $i$ and $j$ and
${\alpha}_{n}$ corresponds to the pole describing the emergence of the bound state.
From the bound state mass spectrum, it can be seen \cite{zamsg} that the soliton amplitude poles appearing on imaginary axis of the $\theta$-plane in the physical region $0<\operatorname{Im}\theta <\pi$  at
\begin{equation}
	\label{alph}
	{\theta}_{n}=i\pi \left( 1 - \frac {n}{\nu} \right) \equiv {\alpha}_{n};\qquad n=1,2,\dots\,,\nu
\end{equation}
correspond to the bound states of the model.

Referring to this method, we first reproduce integral representations of the main spectrum amplitudes. Then, we use these results to obtain our integral representations of scattering amplitudes involving bound states via bootstrap. Our further calculations are based on the use of the formula (\ref{bootdiag}) with the corresponding resonance angles (\ref{alph}).

\medskip

\subsection{Integral representation of the soliton-soliton amplitude}

For later convenience, let us denote the main spectrum S-matrix elements as
\begin{equation}
	\label{sol_sg21}
	\begin{aligned}
		&\\
		{S}_{SG}\left( \theta  \right)=\rho \left( \theta  \right)\left( \begin{matrix}
			a\left( \theta  \right) & 0 & 0 & 0  \\
			0 & b\left( \theta  \right) & c\left( \theta  \right) & 0  \\
			0 & c\left( \theta  \right) & b\left( \theta  \right) & 0  \\
			0 & 0 & 0 & a\left( \theta  \right)  \\
		\end{matrix} \right),\\
	\end{aligned}
\end{equation}
where
\begin{equation}
	\label{sol_sg22}
	\begin{aligned}
		a\left( \theta  \right) &=1, \\ 
		b\left( \theta  \right) &=-\frac{\sin \left( i\nu \theta  \right)}{\sin \left( i\nu \theta +\nu \pi  \right)}, \\ 
		c\left( \theta  \right) &=\frac{\sin \left( \nu \pi  \right)}{\sin \left( i\nu \theta +\nu \pi  \right)}. \\ 
	\end{aligned}
\end{equation}\\
Then, unitarity and crossing symmetry relations for the soliton amplitude $\rho\left(\theta\right)$ take the form
\begin{equation}
	\begin{aligned}
		& \left\{ \begin{aligned}
			\rho \left( \theta  \right)\rho \left( -\theta  \right) &=1 \\ 
			\rho \left( i\pi -\theta  \right) &=-\rho \left( \theta  \right)\frac{\sin \left( i\nu \theta  \right)}{\sin \left( i\nu \theta +\nu \pi  \right)} \\ 
		\end{aligned}\right. \\ 
	\end{aligned}
\end{equation}
Using these relations, as well as the formula \cite{intsys}
\begin{equation}
	\label{ln_int}
	\ln \frac{\sin {\left( \pi x-i\theta  \right)}/{2}\;}{\sin {\left( \pi x+i\theta  \right)}/{2}\;}=2\int_{0}^{\infty }{\frac{\sinh \left( 1-x \right)}{t \sinh t}\sinh \left( \frac{t\theta }{i\pi } \right)} dt,
\end{equation}
it is possible to reproduce the following integral representation of $\rho\left(\theta\right)$ 
\begin{equation}
	\label{rho}
	\rho \left( \theta  \right)=-\exp \int_{0}^{\infty }{\frac{\sinh \left( 1+\nu  \right)t}{t \cosh \nu t\,\sinh t}\sinh \left( \frac{2\nu \theta t}{i\pi } \right)} dt.
\end{equation}
Here the overall minus sign is conventional \cite{bomb}. This representation is particularly useful for further computation of the total sine-Gordon S-matrix using the bootstrap method.

\medskip

\subsection{Integral representations of the first bound state amplitudes}
We provide the derivation of existing integral representations of the soliton-first bound state and first bound state-bound state amplitudes of the sine-Gordon and also perform consistency checks with the trigonometric form of these amplitudes.
This procedure is developed further in the next section, where we obtain our integral representations of amplitudes for more general scattering processes using the same method.

Now we derive the following integral representation of an amplitude ${{S}_{s{{b}_{1}}}}$ for the process $A+{{B}_{1}}\to A+{{B}_{1}}$, corresponding to the soliton-first bound state scattering
\begin{equation}
	\label{isb1}
	{{S}_{s{{b}_{1}}}}\left( \theta  \right)=-\exp 2\int_{0}^{\infty }{\frac{\sinh \left( 1+\nu  \right)t}{t \cosh \nu t}}\cosh \left( \nu t\left( 1-\frac{2\theta }{i\pi } \right) \right) dt.
\end{equation}
Using the integral representation of $\rho \left( \theta \right)$ (\ref{rho}), we get the bootstrap equation (\ref{bootdiag}) for ${{S}_{s{{b}_{1}}}}$ 
\begin{equation}
	\label{bsb1}
	{{S}_{s{{b}_{1}}}}\left( \theta  \right)=S\left( \theta -\frac{i\pi }{2}\left( 1-\frac{1}{\nu } \right) \right){{S}_{T}}\left( \theta +\frac{i\pi }{2}\left( 1-\frac{1}{\nu } \right) \right).
\end{equation}
Denoting
\begin{equation}
	S \left( \theta \right) \equiv \rho \left( \theta \right),\qquad {{S}_{T}}\left( \theta  \right) \equiv b\left( \theta  \right)\rho \left( \theta  \right),
\end{equation}
we transform $b \left( \theta \right)$ into its integral form, using (\ref{ln_int})
\begin{equation}
	\label{b}
	\begin{aligned}
		& \ln b\left( \theta +\frac{i\pi }{2}\left( 1-\frac{1}{\nu } \right) \right)=2\int_{0}^{\infty }{\frac{\sinh \left( 1+\nu  \right)t}{t \sinh t}\sinh \left(t- \frac{2\nu \theta t}{i\pi }\right)} dt \\ 
	\end{aligned}
\end{equation}
and substitute the integral expressions for (\ref{rho}, \ref{b}) into the logarithm of (\ref{bsb1}) to obtain 
\setlength{\jot}{4pt}
\begin{multline}
	\ln S\left( \theta -\frac{i\pi }{2}\left( 1-\frac{1}{\nu } \right) \right){{S}_{T}}\left( \theta +\frac{i\pi }{2}\left( 1-\frac{1}{\nu } \right) \right)
	\\
	= \int_{0}^{\infty }{\frac{\sinh \left( 1+\nu  \right)t}{t \cosh \nu t\sinh t}\left[ -\sinh \left( \nu t-t-\frac{2\nu \theta t}{i\pi } \right)+\sinh \left( \nu t+t-\frac{2\nu \theta t}{i\pi } \right) \right]} dt  
\end{multline}
\setlength{\jot}{2pt}
and
\begin{equation}
	{{S}_{s{{b}_{1}}}}\left( \theta  \right)=-\exp 2\int_{0}^{\infty }{\frac{\sinh \left( 1+\nu  \right)t}{t \cosh \nu t}}\cosh \left( \nu t\left( 1-\frac{2\theta }{i\pi } \right) \right) dt. 
\end{equation}
Bringing this representation to the form
\begin{equation}
	{{S}_{s{{b}_{1}}}} =-\exp 2 \int_{0}^{\infty }{\frac{\sinh \left( 1+\frac{1}{\nu } \right)\frac{p}{2}\sinh \frac{p}{2}}{p \sinh p}}\left( {{e}^{\frac{p}{2}-\frac{p\theta }{i\pi }}}+{{e}^{-\frac{p}{2}+\frac{p\theta }{i\pi }}} \right)dp,
\end{equation}
where $p=2\nu t$,
we can use the relatons \cite{ruijs}
\begin{equation}
	\label{lng_int}
	\ln \frac{\Gamma \left( s+{\left( 1+\sigma +\tau  \right)}/{2}\; \right)\Gamma \left( -s+{\left( 1-\sigma +\tau  \right)}/{2}\; \right)}{\Gamma \left( -s+{\left( 1+\sigma +\tau  \right)}/{2}\; \right)\Gamma \left( s+{\left( 1-\sigma +\tau  \right)}/{2}\; \right)}=2\int\limits_{0}^{\infty }\frac{\sinh \sigma t\sinh 2st}{t \sinh t}{{e}^{-\tau t}}dt
\end{equation}
and \cite{abr}
\begin{equation}
	\label{gcosh}
	\Gamma \left( \frac{1}{2}+iy \right)\Gamma \left( \frac{1}{2}-iy \right)={{\left| \Gamma \left( \frac{1}{2}+iy \right) \right|}^{2}}=\frac{\pi }{\cosh \pi y}
\end{equation}
to write
\setlength{\jot}{6pt}
\begin{multline}
	\ln \left( {{S}_{s{{b}_{1}}}}\left( \theta  \right) \right)\\
	=\ln \frac{\Gamma \left( \frac{3}{4}+\frac{1}{4\nu }+\frac{\theta }{2i\pi } \right)\Gamma \left( -\frac{1}{4}-\frac{1}{4\nu }+\frac{\theta }{2i\pi } \right)}{\Gamma \left( \frac{1}{4}+\frac{1}{4\nu }+\frac{\theta }{2i\pi } \right)\Gamma \left( \frac{1}{4}-\frac{1}{4\nu }+\frac{\theta }{2i\pi } \right)}\frac{\Gamma \left( \frac{5}{4}+\frac{1}{4\nu }-\frac{\theta }{2i\pi } \right)\Gamma \left( \frac{1}{4}-\frac{1}{4\nu }-\frac{\theta }{2i\pi } \right)}{\Gamma \left( \frac{3}{4}+\frac{1}{4\nu }-\frac{\theta }{2i\pi } \right)\Gamma \left( \frac{3}{4}-\frac{1}{4\nu }-\frac{\theta }{2i\pi } \right)} \\
	=\ln \frac{\cosh \left( -i\pi \left( \frac{1}{4}-\frac{1}{4\nu }-\frac{\theta }{2i\pi } \right) \right)\cosh \left( -i\pi \left( \frac{1}{4}+\frac{1}{4\nu }-\frac{\theta }{2i\pi } \right) \right)}{\cosh \left( -i\pi \left( \frac{1}{4}+\frac{1}{4\nu }+\frac{\theta }{2i\pi } \right) \right)\cosh \left( -i\pi \left( \frac{3}{4}+\frac{1}{4\nu }-\frac{\theta }{2i\pi } \right) \right)}\\
	=\ln \frac{\sinh \theta +i\sin \left( \frac{\pi }{2\nu }+\frac{\pi }{2} \right)}{\sinh \theta -i\sin \left( \frac{\pi }{2\nu }+\frac{\pi }{2} \right)}.
\end{multline}
\setlength{\jot}{2pt}
Thus, we have shown that the integral representation for the soliton-first bound state amplitude (\ref{isb1}) is equivalent to the trigonometric form (\ref{zamsbn})
\begin{equation}
	{{S}_{s{{b}_{1}}}}\left( \theta  \right)=\frac{\sinh \theta +i\sin \frac{\pi }{2}\left( 1+\frac{1}{\nu } \right)}{\sinh \theta -i\sin \frac{\pi }{2}\left( 1+\frac{1}{\nu } \right)}.
\end{equation}
Next, we derive the following integral representation of an amplitude ${{S}_{{{b}_{1}}{{b}_{1}}}}$ for the process ${{B}_{1}}+{{B}_{1}}\to {{B}_{1}}+{{B}_{1}} $, corresponding to the first bound state-first bound state scattering.
\begin{equation}
	\label{ib1b1}
	{{S}_{{{b}_{1}}{{b}_{1}}}}\left( \theta  \right)=-\exp \int_{0}^{\infty }{4\frac{\sinh \left( 1+\nu  \right)t}{t \cosh \nu t}}\cosh \left( \nu t\left( 1-\frac{2\theta }{i\pi } \right) \right)\cosh \left( \nu t-t \right) dt.
\end{equation}
As in the previous case, we start from the booststrap equation (\ref{bootdiag})
\begin{equation}
	\label{bb1b1}
	{{S}_{{{b}_{1}}{{b}_{1}}}}\left( \theta  \right)={{S}_{s{{b}_{1}}}}\left( \theta -\frac{i\pi }{2}\left( 1-\frac{1}{\nu } \right) \right){{S}_{s{{b}_{1}}}}\left( \theta +\frac{i\pi }{2}\left( 1-\frac{1}{\nu } \right) \right).
\end{equation}
Using the integral representation of ${S}_{{b}_{1}}$ (\ref{isb1}) together with the formula (\ref{ln_int}), the bootsrap equation can be transformed into the stated result (\ref{ib1b1}).

Repeating the procedure for the soliton-first bound state case leads to the known trigonometric form (\ref{zamsbn}) of the integral amplitude (\ref{ib1b1})
\begin{equation}
	\begin{aligned}
		{{S}_{{{b}_{1}}{{b}_{1}}}}\left( \theta  \right)=\frac{\sinh \theta +i\sin \frac{\pi }{\nu }}{\sinh \theta -i\sin \frac{\pi }{\nu }}.
	\end{aligned}
\end{equation}
Therefore, the results obtained via boostrap are consistent with other existing representations and we can proceed to more general calculations of this kind.

\medskip
\subsection{Integral representations of the general bound state amplitudes}

Now we derive the scattering amplitudes for bound states of the sine-Gordon model in their integral form using the bootstrap method.

We have expressed the amplitude for the process ${A+{{B}_{n}}\to A+{{B}_{n}}}$ describing the soliton - $n$-th bound state scattering in the following integral form:
\begin{equation}
	\label{sbn}
	\begin{aligned}
		&{{S}_{s{{b}_{n}}}}\left( \theta  \right)=-\exp \int_{0}^{\infty }{4\frac{\sinh \left( 1+\nu  \right)t\sinh \nu t\sinh nt}{t\sinh 2\nu t\sinh t}}\cosh \left( \nu t\left( 1-\frac{2\theta }{i\pi } \right) \right) dt.\\
	\end{aligned}	
\end{equation}
Inserting the pole corresponding to the $n$-th bound state into the bootstrap formula (\ref{bootdiag}), we write
\begin{equation}
	\label{bsbn}
	{{S}_{s{{b}_{n}}}}\left( \theta  \right)=S\left( \theta -\frac{i\pi }{2}\left( 1-\frac{n}{\nu } \right) \right){{S}_{T}}\left( \theta +\frac{i\pi }{2}\left( 1-\frac{n}{\nu } \right) \right)
\end{equation}
From now on, we return to the notation (\ref{sol_sg22}): 
$S \left( \theta \right) \equiv \rho \left( \theta \right), \quad {{S}_{T}}\left( \theta  \right) \equiv b\left( \theta  \right)\rho \left( \theta  \right) $ and use integral representations (\ref{rho}, \ref{b}).
Then, for the logarithm of (\ref{bsbn}) we have
\setlength{\jot}{6pt}
\begin{multline}
	\ln S\left( \theta -\frac{i\pi }{2}\left( 1-\frac{n}{\nu } \right) \right){{S}_{T}}\left( \theta +\frac{i\pi }{2}\left( 1-\frac{n}{\nu } \right) \right)\\ 
	\shoveleft{= \int_{0}^{\infty }{\frac{\sinh \left( 1+\nu  \right)t}{t \cosh \nu t\sinh t}} \sinh \left( \frac{2\nu t}{i\pi }\left( \theta +\frac{i\pi }{2}\left( 1-\frac{n}{\nu } \right) \right) \right)dt}\\ 
	\shoveleft{+ \int_{0}^{\infty }{\frac{\sinh \left( 1+\nu  \right)t}{t \cosh \nu t\sinh t}} \sinh \left( \frac{2\nu t}{i\pi }\left( \theta -\frac{i\pi }{2}\left( 1-\frac{n}{\nu } \right) \right) \right) }dt\\ 
	\shoveleft{+\int_{0}^{\infty }{2\frac{\sinh \left( 1+\nu  \right)t}{t\sinh t}\sinh \left( nt-\frac{2\nu \theta t}{i\pi } \right)}dt}\\ 
	=\int_{0}^{\infty }{4\frac{\sinh \left( 1+\nu  \right)t\sinh \nu t\sinh nt}{t \sinh 2\nu t\sinh t}}\cosh \left( \nu t\left( 1-\frac{2\theta }{i\pi } \right) \right)dt. 
\end{multline}
\setlength{\jot}{2pt}
Thus, we obtain the representation (\ref{sbn}).

Considering the ${{B}_{m}}+{{B}_{n}}\to {{B}_{m}}+{{B}_{n}}$ process of bound state-bound state scattering, we have obtained the following integral representation of the corresponding amplitude
\begin{multline}
	\label{bmbn}
	{{S}_{{{b}_{m}}{{b}_{n}}}}\left( \theta  \right)\\
	=-\exp \int_{0}^{\infty }{8\frac{\sinh \left( 1+\nu  \right)t\sinh \nu t\sinh mt\cosh \left( \nu t-nt \right)}{t \sinh 2\nu t\sinh t}}\cosh \left( \nu t-\frac{2\nu t\theta }{i\pi } \right)dt.
\end{multline}
The starting boostrap equation (\ref{bootdiag}) is
\begin{equation}
	\label{bbmbn}
	{{S}_{{{b}_{m}}{{b}_{n}}}}\left( \theta  \right)={{S}_{s{{b}_{m}}}}\left( \theta -\frac{i\pi }{2}\left( 1-\frac{n}{\nu } \right) \right){{S}_{s{{b}_{m}}}}\left( \theta +\frac{i\pi }{2}\left( 1-\frac{n}{\nu } \right) \right).
\end{equation}
We then transform its logarithm
\setlength{\jot}{6pt}
\begin{multline}
	\ln \left( -{{S}_{{{b}_{m}}{{b}_{n}}}}\left( \theta  \right) \right)\\
	=\int_{0}^{\infty }{4\frac{\sinh \left( 1+\nu  \right)t\sinh \nu t\sinh mt}{t\sinh 2\nu t\sinh t}}
	\cosh \left( \nu t-\frac{2\nu t}{i\pi }\left( \theta -\frac{i\pi }{2}\left( 1-\frac{n}{\nu } \right) \right) \right)dt\\
	+\int_{0}^{\infty }{4\frac{\sinh \left( 1+\nu  \right)t\sinh \nu t\sinh mt}{t\sinh 2\nu t\sinh t}} 
	\cosh \left( \nu t-\frac{2\nu t}{i\pi }\left( \theta +\frac{i\pi }{2}\left( 1-\frac{n}{\nu } \right) \right) \right) dt\\ 
	=\int_{0}^{\infty }{8\frac{\sinh \left( 1+\nu  \right)t\sinh \nu t\sinh mt\cosh \left( \nu t-nt \right)}{t \sinh 2\nu t\sinh t}}\cosh \left( \nu t-\frac{2\nu t\theta }{i\pi } \right)dt.
\end{multline}
\setlength{\jot}{2pt}
and arrive at the expression (\ref{bmbn}).

We are aware that some other integral representations of the sine-Gordon amplitudes involving bound states appear in the works on this subject. Nevertheless, we believe that our results can be useful for studying analytical properties of the amplitudes and for further investigation of the connection between amplitudes and relevant form factors.

We have also succeeded in performing consistency checks with the trignonmetric representations (\ref{zamsbn}) and (\ref{zambmbn}) for the cases of $A+{{B}_{n}}\to A+{{B}_{n}}$ process with $n=1,\,2,\,3,\,4$, and of ${{B}_{m}}+{{B}_{n}}\to {{B}_{m}}+{{B}_{n}}$ process with $m=1,\,2,\,3,\,4$ and $n$ arbitrary using the same method as described in the previous section.

For example, for the process $A+{{B}_{4}}\to A+{{B}_{4}}$ we get
\setlength{\jot}{8pt}
\begin{multline}
	\label{lnsb4}
	\ln \left( -{{S}_{s{{b}_{4}}}}\left( \theta  \right) \right)=\int_{0}^{\infty }{4\frac{dt}{t}\frac{\sinh \left( 1+\nu  \right)t\sinh \nu t\sinh 4t}{\sinh 2\nu t\sinh t}}\cosh \left( \nu t\left( 1-\frac{2\theta }{i\pi } \right) \right)\\
	=\ln \left(-\frac{\sinh \theta +i\cos \frac{2\pi }{\nu }}{\sinh \theta -i\cos \frac{2\pi }{\nu }}\frac{{{\sin }^{2}}\left( -\frac{\pi }{2\nu }-\frac{\pi }{4}+\frac{i\theta }{2} \right)}{{{\sin }^{2}}\left( -\frac{\pi }{2\nu }-\frac{\pi }{4}-\frac{i\theta }{2} \right)}\right)\\
	+\ln \frac{{{\sin }^{2}}\left( -\frac{\pi }{4}+\frac{i\theta }{2} \right){{\sin }^{2}}\left( \frac{\pi }{2\nu }-\frac{\pi }{4}+\frac{i\theta }{2} \right)}{{{\sin }^{2}}\left( -\frac{\pi }{4}-\frac{i\theta }{2} \right){{\sin }^{2}}\left( \frac{\pi }{2\nu }-\frac{\pi }{4}-\frac{i\theta }{2} \right)},
\end{multline}
\setlength{\jot}{2pt}
and
\begin{multline}
	\label{sb4}
	{{S}_{s{{b}_{4}}}}\left( \theta  \right)\\
	=\frac{\sinh \theta +i\cos \frac{2\pi }{\nu }}{\sinh \theta -i\cos \frac{2\pi }{\nu }}\frac{{{\sin }^{2}}\left( -\frac{\pi }{2\nu }-\frac{\pi }{4}+\frac{i\theta }{2} \right){{\sin }^{2}}\left( -\frac{\pi }{4}+\frac{i\theta }{2} \right){{\sin }^{2}}\left( \frac{\pi }{2\nu }-\frac{\pi }{4}+\frac{i\theta }{2} \right)}{{{\sin }^{2}}\left( -\frac{\pi }{2\nu }-\frac{\pi }{4}-\frac{i\theta }{2} \right){{\sin }^{2}}\left( -\frac{\pi }{4}-\frac{i\theta }{2} \right){{\sin }^{2}}\left( \frac{\pi }{2\nu }-\frac{\pi }{4}-\frac{i\theta }{2} \right)}. \\ 
\end{multline}
For ${{B}_{4}}+{{B}_{n}}\to {{B}_{4}}+{{B}_{n}}$ scattering we have
\setlength{\jot}{6pt}
\begin{multline}
	\label{b4bn}
	\ln \left( -{{S}^{\left( 4,n \right)}}\left( \theta  \right) \right)
	=\int_{0}^{\infty }{4\frac{\sinh \left( 1+\nu  \right)t\sinh \nu t\sinh 4t}{t \sinh 2\nu t\sinh t}}\cosh \left( -\frac{2\nu t\theta }{i\pi }+nt \right) dt\\
	+\int_{0}^{\infty }{4\frac{\sinh \left( 1+\nu  \right)t\sinh \nu t\sinh 4t}{t \sinh 2\nu t\sinh t}} \cosh \left( 2\nu t-\frac{2\nu t\theta }{i\pi }-nt \right)dt,
\end{multline}
so that the amplitude takes the form
\begin{multline}
	{{S}^{\left( 4,n \right)}}\left( \theta  \right) =\frac{\sinh \theta +i\sin \frac{n+4}{2\nu }}{\sinh \theta -i\sin \frac{n+4}{2\nu }}\frac{\sinh \theta +i\sin \frac{n-4}{2\nu }}{\sinh \theta -i\sin \frac{n-4}{2\nu }}\\ 
	\times \frac{{{\sin }^{2}}\left( \frac{-n+2}{4\nu }+\frac{i\theta }{2} \right){{\sin }^{2}}\left( \frac{-n}{4\nu }+\frac{i\theta }{2} \right){{\sin }^{2}}\left( \frac{-n-2}{4\nu }+\frac{i\theta }{2} \right)}{{{\sin }^{2}}\left( \frac{-n+2}{4\nu }-\frac{i\theta }{2} \right){{\sin }^{2}}\left( \frac{-n}{4\nu }-\frac{i\theta }{2} \right){{\sin }^{2}}\left( \frac{-n-2}{4\nu }-\frac{i\theta }{2} \right)} \\
	\times \frac{{{\cos }^{2}}\left( \frac{n-2}{4\nu }+\frac{i\theta }{2} \right){{\cos }^{2}}\left( \frac{n}{4\nu }+\frac{i\theta }{2} \right){{\cos }^{2}}\left( \frac{n+2}{4\nu }+\frac{i\theta }{2} \right)}{{{\cos }^{2}}\left( \frac{n-2}{4\nu }-\frac{i\theta }{2} \right){{\cos }^{2}}\left( \frac{n}{4\nu }-\frac{i\theta }{2} \right){{\cos }^{2}}\left( \frac{n+2}{4\nu }-\frac{i\theta }{2} \right)}. 
\end{multline}
\setlength{\jot}{2pt}
Thus, our results (\ref{sbn}) and (\ref{bmbn}) for the sine-Gordon scattering involving bound states are supposed to coincide with the trigonometric solutions (\ref{zamsbn}) and (\ref{zambmbn}), respectively.

\bigskip

\section{Correlation functions from the S-matrix: form factors and celestial amplitudes}

In this section we discuss the correspondence between scattering amplitudes and correlation functions. In particular, we consider form factors defined as matrix elements of operators, and celestial amplitudes that are conformal correlators obtained by mapping scattering amplitudes to conformal basis. We use our previous results for the integral sine-Gordon amplitudes to compute minimal form factors of the sine-Gordon model. Then, we review the latest developements in the bootstrap method in the context of integrable two-dimensional amplitudes and celestial amplitudes. We also obtain analytical constraints by transforming the general properties of the sine-Gordon S-matrix and the bootstrap equation to the Fourier space of celestial amplitudes.

\subsection{Minimal form factors of the sine-Gordon model}

The functions introduced in \cite{sff}
\begin{equation}
	\label{ffdef}
	\begin{aligned}
		&F_{n}^{{\mathcal{O}}}\left( {{\theta }_{1}},{{\theta }_{2}},\ldots ,{{\theta }_{n}} \right)=\left\langle  0\, \right|{\mathcal{O}}\left( 0 \right){{\left| {{\theta }_{1}},{{\theta }_{2}},\ldots ,{{\theta }_{n}} \right\rangle }_{in}}\\
	\end{aligned}
\end{equation}

are called form factors, namely the matrix elements of a local scalar hermitian operator ${\mathcal{O}}\left( x \right)$ placed between an $n$-particle state and vacuum. Form factors are used for calculating correlation function in various models. Here we use the notation
$\left| \ldots A\left( {{\theta }_{n}} \right)\ldots  \right\rangle \equiv \left| \ldots {{\theta }_{n}}\ldots  \right\rangle $.

Form factors can be treated as functions of rapidity differences ${{\theta }_{ij}}$ due to Lorentz invariance
\begin{equation}	
	\begin{aligned}
		&F_{n}^{{\mathcal{O}}}\left( {{\theta }_{1}},{{\theta }_{2}},\ldots ,{{\theta }_{n}} \right)=F_{n}^{{\mathcal{O}}}\left( {{\theta }_{12}},{{\theta }_{13}},\ldots ,{{\theta }_{ij}},\ldots  \right),\, i<j.\\
	\end{aligned}
\end{equation}
Form factors $F_{n}^{\mathcal{O}} $ are analytical functions in the physical region $0 <\operatorname{Im}\theta_{ij} < 2\pi$, except for the poles corresponding to bound states.

General matrix elements are constructed in the following way due to crossing invariance
\begin{equation}
	F_{n+m}^{O}\left( {{\theta }_{1}},{{\theta }_{2}},\ldots ,{{\theta }_{m}},{{\theta }_{m+1}}-i\pi ,\ldots ,{{\theta }_{n}}-i\pi  \right)= F_{n+m}^{O}\left( {{\theta }_{ij}},i\pi -{{\theta }_{sr}},{{\theta }_{kl}} \right),
\end{equation}
where $1\le i <j \le m,\quad1\le r \le m <s \le n,\quad m<k<l\le n$.

Form factors satisfy the so-called Watson equations
\begin{equation}
	\label{wats}
	\begin{aligned}
		F_{n}^{\mathcal{O}}\left( {{\theta }_{1}},\ldots ,{{\theta }_{i}},{{\theta }_{i+1}},\ldots {{\theta }_{n}} \right) &=F_{n}^{\mathcal{O}}\left( {{\theta }_{1}},\ldots {{\theta }_{i+1}},{{\theta }_{i}},\ldots {{\theta }_{n}} \right)S\left( {{\theta }_{i}}-{{\theta }_{i+1}} \right), \\ 
		F_{n}^{\mathcal{O}}\left( {{\theta }_{1}}+2\pi i,\ldots ,{{\theta }_{n-1}},{{\theta }_{n}} \right) &=\prod\limits_{i=2}^{n}{S\left( {{\theta }_{i}}-{{\theta }_{1}} \right)}F_{n}^{\mathcal{O}}\left( {{\theta }_{1}},\ldots ,{{\theta }_{n}} \right), \\ 
	\end{aligned}
\end{equation}
where the first equation can be treated as a consequence of interpreting permutations of two operators as scattering processes.

General solution of the Watson equations is written as
\begin{equation}
	\label{watssol}
	\begin{aligned}
		&F_{n}^{{\mathcal{O}}}\left( {{\theta }_{1}},{{\theta }_{2}},\ldots ,{{\theta }_{n}} \right)=K_{n}^{{\mathcal{O}}}\left( {{\theta }_{1}},\ldots ,{{\theta }_{n}} \right)\prod\limits_{i<j}{{{F}^{\min }}\left( {{\theta }_{ij}} \right)},\\
	\end{aligned}
\end{equation}
where minimal two-particle form factors ${{F}^{\min }}\left( \theta  \right)$ are analytic functions in the region $0\le \operatorname{Im}\theta  \le \pi$ that solve the given system of equations and contain no poles or zeros in $0<\operatorname{Im}\theta <\pi$.

The Watson equations simplify for the case of $n=2$:
\begin{equation}
	\label{wats2}
	\begin{aligned}
		F_{2}^{{\mathcal{O}}}\left( \theta \right) &=F_{2}^{{\mathcal{O}}}\left( -\theta  \right){{S}_{2}}\left( \theta  \right),  \\
		F_{2}^{{\mathcal{O}}}\left( i\pi -\theta \right)&=F_{2}^{{\mathcal{O}}}\left( i\pi +\theta \right) , \\
	\end{aligned}
\end{equation}
and the general solution in this case has the form
\begin{equation}
	\label{watssol2}
	\begin{aligned}
		F\left( \theta  \right)&={K}_{2}\left( \theta  \right){{F}^{\min }}\left( \theta  \right), \\ 
		{K}_{2}\left( \theta  \right)&={K}_{2}\left( -\theta  \right)={K}_{2}\left( 2\pi i+\theta  \right). \\ 
	\end{aligned}
\end{equation}
The function satisfying these equations is
\begin{equation}
	\label{ksol}
	{K}_{2}\left( \theta  \right)=const\prod\nolimits_{k=1}^{L}{{{\left[ \sinh\frac{1}{2}\left( \theta -i{{\alpha }_{k}} \right)\sinh\frac{1}{2}\left( \theta +i{{\alpha }_{k}} \right) \right]}^{-1}}},
\end{equation}
where the values of $\theta ={{\alpha }_{k}}$ correspond to the poles.

There is an important relation between the S-matrix integral representations and form factors \cite{kw}. When the S-matrix can be represented as
\begin{equation}
	\label{sint}
	S\left( \theta  \right)=\exp \int_{0}^{\infty }{\frac{f\left( t \right)}{t}\sinh \frac{t\theta }{i\pi }dt},
\end{equation}
the corresponding form factor takes the form
\begin{equation}
	\label{fint}
	{{F}^{\min }}\left( \theta  \right)\propto\exp \int_{0}^{\infty }{\frac{f\left( t \right)}{t\sinh t}{{\sin }^{2}}\left( \frac{t\left( i\pi -\theta  \right)}{2\pi} \right)dt}.
\end{equation}

Now we demonstrate the application of obtained results for the sine-Gordon scattering amplitudes to the computation of some minimal form factors in this model. The integral representation of the soliton-soliton scattering amplitude (\ref{rho}) leads to the following minimal soliton-soliton form factor
${ F_{ss}^{SG\min }}\left( \theta,\,\nu  \right)$
\begin{equation} 
	\label{fssmin}
	\begin{aligned}
		&{ F_{ss}^{SG\min }}\left( \theta,\nu  \right)\propto\exp \int_{0}^{\infty }{\frac{\sinh \left( \frac{1}{\nu }+1 \right)\frac{t}{2}}{t \cosh \frac{t}{2}\,\sinh \frac{t}{2\nu }\sinh t}{{\sin }^{2}}\left( \frac{t\left( i\pi -\theta  \right)}{2\pi} \right)dt}.
	\end{aligned}
\end{equation} 
For the first sine-Gordon bound states, the minimal form factor is
\begin{equation}
	\label{fb1b1}
	F_{{{b}_{1}}{{b}_{1}}}^{SG\min }\left( \theta,\,\nu  \right)\propto\exp2\int_{0}^{\infty }{\frac{\cosh \left( \frac{t}{2}-\frac{t}{\nu } \right)}{t \cosh \frac{t}{2}\sinh t}{{\sin }^{2}}\frac{t\left( i\pi -\theta  \right)}{2\pi }dt}
\end{equation}
since the corresponding amplitude ${{S}_{{b}_{1}{b}_{1}}} \left( \theta \right) $ (\ref{ib1b1}) can be brought to the form 
\setlength{\jot}{2pt}
\begin{equation}
	{{S}_{{{b}_{1}}{{b}_{1}}}}\left( \theta  \right) =-\exp 2\int_{0}^{\infty }{\frac{\cosh \left( \frac{t}{2}-\frac{t}{\nu } \right)}{t\cosh \frac{t}{2}}\sinh \frac{t\theta }{i\pi }dt}
\end{equation}
\setlength{\jot}{2pt}
using the technique described in the previous section. The form factor (\ref{fb1b1}) is obtained directly using (\ref{fint}).

We suggest further applications of the results obtained in the previous section to calculation of form factors of a more general kind. Investigating the connection between scattering amplitudes and form factors could possibly lead to more general correspondence rules than the one considered above.

\subsection{Two-dimensional integrable S-matrix in conformal basis}
One of the important questions that has led to the study of scattering amplitudes in the so-called conformal basis is the search for hidden symmetries of the amplitudes. Some of the other bases developed for this purpose include twistor basis \cite{twist_sym} and closely related spinor helicity basis \cite{spin_sym}, that illustrate the symmetries of gauge theory amplitudes and gravity amplitudes.

The problem of flat space holography \cite{hol}, namely the search for two-dimensional conformal symmetries in four-dimensional amplitudes, has lead to the discovery of the conformal basis of wavefunctions, where scattering amplitudes are interpreted as $d$-dimensional conformal correlators, which are called celestial amplitudes. This possibility is illustrated by the fact that the Lorentz group in $\mathbb{R}^{1,\: d+1}$ is identical to the Euclidean conformal group in $d$ dimensions. Recently this approach has been developed for integrable two-dimensional amplitudes, where significant simplifications in the expression for celestial amplitudes take place. Namely, the celestial amplitude becomes a Fourier transform of a rapidity-parametrized 2$d$-integrable S-matrix, and the bootstrap approach can be further formulated in Fourier space.

Let us briefly describe the general construction of celestial amplitudes \cite{conf_bas} in order to illustrate the simplification for the 2$d$-integrable case.

The massive scalar conformal primary wavefunction $\phi_\Delta\left(X^\mu;\vec{\omega}\right)$ of mass $m$ in $\mathbb{R}^{1,\: d+1}$ is a wavefunction labelled by a conformal dimension $\Delta$ and a point $\vec{\omega}$ in $\mathbb{R}^d$. The Fourier expansion of the conformal primary wavefunction is
\begin{equation}
	\label{prim_exp}
	\phi_{\Delta}^{\pm}\left(X^\mu;\vec{\omega}\right)=\int_{H_{d+1}}G_{\Delta}\left(\hat{p};\vec{\omega}\right)\exp\left[\pm im\hat{p}\cdot X\right]\left[d\hat{p}\right],
\end{equation}
where the on-shell momenta, a unit timelike vector $\hat{p}\left(y,\vec{z}\right)$ satisfying $\hat{p}^2=-1$
\begin{equation} \hat{p}^\mu_i\equiv\hat{p}^\mu\left(\theta_i\right)=\frac{p_i}{m}=\left(\cosh\theta_i,\;\sinh\theta_i\right)
\end{equation}
can be parametrized using the coordinates $y,\;\vec{z}$ on $H_{d+1}$ with $y>0$ and $\vec{z}\in \mathbb{R}^d$ as
\begin{equation}
	\label{moment_par}	
	\hat{p}\left(y,\vec{z}\right)=\left(\frac{1+y^2+\vert\vec{z}\vert^2}{2y}, \frac{\vec{z}}{y},\frac{1-y^2-\vert\vec{z}\vert^2}{2y}\right),
\end{equation}
and $\left[d\hat{p}\right]$ is the $SO\left(1,d+1\right)$ invariant measure on $H_{d+1}$
\begin{equation}
	\label{meas}
	\int_{H_{d+1}}\left[d\hat{p}\right]\equiv\int_{0}^\infty \frac{dy}{y^{d+1}}\int d^d\vec{z}=\int\frac{d^{d+1}\hat{p}^i}{\hat{p}^0},\quad i=1,2,\dots,d+1,\quad \hat{p}^0=\sqrt{\hat{p}^i \hat{p}^i +1}.
\end{equation}
The scalar bulk-to-boundary propagator $G_{\Delta}\left(\hat{p};\vec{\omega}\right)$ in $H_{d+1}$ is given by
\begin{equation}
	\label{prop_1}
	G_{\Delta}\left(\hat{p};\vec{\omega}\right)=\left(\frac{y}{y^2+|\vec{z}-\vec{\omega}|^2}\right)^{\Delta},
\end{equation}
where $\vec{\omega}\in\mathbb{R}^d$ lies on the boundary of $H_{d+1}$. Given a map from $\mathbb{R}^d$ to a unit null momentum $q^\mu$ in $\mathbb{R}^{1,\: d+1}$
\begin{equation}
	\label{null_moment}
	q^{\mu}\left(\vec{\omega}\right)=\left(1+|\vec{\omega}|^2,\; 2\vec{\omega}, \; 1-|\vec{\omega}|^2\right),
\end{equation}
the scalar bulk-to-boundary propagator can be parametrized in terms of $\hat{p}^\mu\left(y,\vec{z}\right)$ and $q^\mu\left(\vec{\omega}\right)$ as
\begin{equation}
	\label{prop_2}
	G_{\Delta}\left(\hat{p};q\right)=\frac{1}{\left(-\hat{p}\cdot q\right)^\Delta}.
\end{equation}
The change of basis (\ref{prim_exp}) can be extended to scattering amplitudes in momentum space $\mathcal{A}\left(p_i^{\mu}\right)$ as an integral transform to the basis of conformal primary wavefunctions
\begin{equation}
	\label{amplit_conf_bas}
	\tilde{\mathcal{A}}\left(\delta_i,\vec{\omega}_i\right)\equiv\prod_{k=1}^{n}\int_{H_{d+1}} G_{\Delta_k}\left(\hat{p}_k ; \vec{\omega}_k\right) \mathcal{A}\left(\pm m_i \hat{p}_i^\mu\right)\left[d\hat{p}_k\right],
\end{equation}
and the quantity $\tilde{\mathcal{A}}\left(\delta_i,\vec{\omega}_i\right)$ is known as the massive celestial amplitude. Due to the conformal invariance of the primary wavefunctions, the celestial amplitude transforms covariantly as a $d$-dimensional CFT $n$-point function.

It has been shown in \cite{cel_2d, kap_trop} that the celestial 4-point amplitude for $2\to2$ scattering of massive scalar particles in $2d$ 
\begin{equation}
	\label{2d_cel}
	\mathcal{A}=\left(\prod_{i=1}^4 \int\frac{d\hat{p}^1_i}{\hat{p}^0_i}\right)\prod^4_{i=1}G_{\Delta_i}\left(\hat{p}_i\right) S_{2\to2},
\end{equation}
where $G_{\Delta_i}\left(\hat{p}_i\right)$ is the bulk-to-boundary propagator in $H_1$,
simplifies to a Fourier transform of the S-matrix with respect to rapidity
\begin{equation}
	\label{fourier_ampl}
	\mathcal{A}\left(\omega\right)=\int_{-\infty}^{\infty} e^{i\omega\theta} S\left(\theta\right)d\theta.
\end{equation}

This expression has been analyzed perturbatively for the sinh-Gordon model, which is the analyctic continuation of the sine-Gordon model given by the change of the parameter $\beta\to i\beta$ \cite{intsys}. The sinh-Gordon S-matrix is the simplest one satisfying the fundamental properties for  $O\left(2\right)$ two-dimensional integrable theories. It takes the form
\begin{equation}
	\label{sinhG}
	S_{shG}\left(\theta\right)=\frac{\sinh\theta-i\sin\alpha}{\sinh\theta+i\sin\alpha},\qquad \alpha=\frac{\pi \beta^2}{8\pi+\beta^2},
\end{equation}
and corresponds to the analytic continuation of the sine-Gordon S-matrix describing to the first bound state-first bound state scattering. Since in this case the S-matrix $S\left(\theta\right)$ contains poles at $\theta=0$, it is convenient to define the retarded and advanced perturbative celestial amplitudes $\mathcal{A}^+\left(\omega\right)$ and $\mathcal{A}^-\left(\omega\right)$ as
\begin{equation}
	\label{fourier_ampl_ie}
	\mathcal{A}^\pm\left(\omega\right)=\int_{-\infty}^{\infty} e^{i\omega\theta} S\left(\theta\pm i\epsilon\right)d\theta.
\end{equation}
In this formalism, the bootstrap principle can be applied to the retarded and advanced celestial amplitudes. In the work \cite{cel_2d}, the crossing condition $S\left(\theta\right)=s\left(i\pi-\theta\right)$ is reformulated as
\begin{equation}
	\label{cross_fourier}
	\mathcal{A}^+\left(\omega\right)=e^{-\omega\pi}\mathcal{A}^+\left(-\omega\right),\qquad\mathcal{A}^-\left(\omega\right)=e^{\omega\pi}\mathcal{A}^-\left(-\omega\right),
\end{equation}
and the unitarity condition for the rapidity-parametrized S-matrix gives in Fourier space
\begin{equation}
	\label{unit_fourier}
	2\pi\delta\left(\omega\right)=\frac{1}{2\pi}\int^{\infty}_{-\infty}\mathcal{A}^+\left(\omega+\omega^\prime\right)\mathcal{A}^{-}\left(\omega^\prime\right)d\omega^\prime.
\end{equation}

The important question that remains open is non-perturbative analysis of the Fourier transform of the two-dimensional integrable S-matrix and the formulation of the bootstrap principle in this case. In the context of conformal field theory, the main purpose of bootstrap approach is to obtain the higher order conformal correlators from the lower order ones.

\subsection{Sine-Gordon analytical constraints in celestial space}

Computing celestial amplitudes corresponding to other integrable 2-dimensional S-matrices is an interesting prospect of further study. Here we reformulate the analytical properties established for the soliton-soliton and soliton-antisoliton scattering in the sine-Gordon model in conformal basis. We also speculate on the possible extension of the general bootstrap formula to the celestial Fourier space.  

The sine-Gordon model possesses a more complicated S-matrix than the simplest sinh-Gordon case. This leads to the non-trivial crossing and unitarity relations reviewed earlier in this work.

The sine-Gordon S-matrix unitarity property takes the form
\begin{equation}
	\label{unit_sg1}
	\begin{aligned}
		S\left( \theta  \right)S\left( -\theta  \right)&=1, \\ 
		{{S}_{T}}\left( \theta  \right){{S}_{T}}\left( -\theta  \right)+{{S}_{R}}\left( \theta  \right){{S}_{R}}\left( -\theta  \right)&=1, \\ 
		{{S}_{T}}\left( \theta  \right){{S}_{R}}\left( -\theta  \right)+{{S}_{R}}\left( \theta  \right){{S}_{T}}\left( -\theta  \right)&=0, \\ 
	\end{aligned}
\end{equation}
and crossing symmetry relation is
\begin{equation}
	\label{cross_sg1}
	\begin{aligned}
		S\left( \theta  \right)&={{S}_{T}}\left( i\pi -\theta  \right),\\ 
		{{S}_{R}}\left( \theta  \right)&={{S}_{R}}\left( i\pi -\theta  \right), \\ 
	\end{aligned}
\end{equation}
where $S\left( \theta \right)$ is the soliton-soliton scattering amplitude, ${S}_{T} \left( \theta \right)$ is the soliton-antisoliton transition amplitude and  ${S}_{R} \left( \theta \right)$ is the soliton-antisoliton reflection amplitude.

Following the general framework for two-dimensional integrable models introduced in \cite{cel_2d}, we define the corresponding celestial amplitudes in conformal basis in accordance with (\ref{fourier_ampl}) as
\setlength{\jot}{6pt}
\begin{equation}
	\label{fourier_sg_ampl}
	\begin{aligned}
		\mathcal{A}\left(\omega\right)&=\int_{-\infty}^{\infty} e^{i\omega\theta} S\left(\theta\right)d\theta,\\
		\mathcal{A}_T\left(\omega\right)&=\int_{-\infty}^{\infty} e^{i\omega\theta} S_T\left(\theta\right)d\theta,\\
		\mathcal{A}_R\left(\omega\right)&=\int_{-\infty}^{\infty} e^{i\omega\theta} S_R\left(\theta\right)d\theta,\\
	\end{aligned}
\end{equation}
where we refer to the notations for the sine-Gordon S-matrix. Corresponding celestial amplitudes with the $i\epsilon$-prescription are assumed to be defined as in (\ref{fourier_ampl_ie}). 

Then, we can expand the analytical conditions imposed on the sine-Gordon S-matrix to the corresponding retarded and advanced celestial amplitudes. In this case, the unitarity conditions are
\begin{equation}
	\label{fourier_unit_sg}
	\begin{aligned}
		2\pi\delta\left(\omega\right)&=\frac{1}{2\pi}\int^{\infty}_{-\infty}\mathcal{A}^+\left(\omega+\omega^\prime\right)\mathcal{A}^{-}\left(\omega^\prime\right)d\omega^\prime,\\
		2\pi\delta\left(\omega\right)&=\frac{1}{2\pi}\int^{\infty}_{-\infty}\left[\mathcal{A}^+_T\left(\omega+\omega^\prime\right)\mathcal{A}^{-}\left(\omega^\prime\right)_T+\mathcal{A}^+_R\left(\omega+\omega^\prime\right)\mathcal{A}^{-}_R\left(\omega^\prime\right)\right]d\omega^\prime,\\
		2\pi\delta\left(\omega\right)&=\frac{1}{2\pi}\int^{\infty}_{-\infty}\left[\mathcal{A}^+_T\left(\omega+\omega^\prime\right)\mathcal{A}^{-}_R\left(\omega^\prime\right)+\mathcal{A}^+_R\left(\omega+\omega^\prime\right)\mathcal{A}^{-}_T\left(\omega^\prime\right)\right]d\omega^\prime,\\
	\end{aligned}
\end{equation}
and crossing symmetry in celestial space is
\begin{equation}
	\label{fourier_cross_sg}
	\begin{aligned}
		\mathcal{A}^\pm\left(\omega\right)&=e^{\mp\omega\pi}\mathcal{A}_T^\pm\left(-\omega\right),\\
		\mathcal{A}^\pm_R\left(\omega\right)&=e^{\mp\omega\pi}\mathcal{A}_R^\pm\left(-\omega\right).\\
	\end{aligned}
\end{equation}

Now we consider the general S-matrix bootstrap for diagonal scattering (\ref{bootdiag}) and treat it as the fusion product of two amplitudes with given arguments. For simplicity, we rewrite the bootstrap equation as
\begin{equation}
	\label{bootdiag_1}
	{{S}_{N}}\left( \theta  \right)={{S}_{I}}\left( \theta -\zeta \right){{S}_{J}}\left( \theta +\zeta \right),
\end{equation}
where ${S}_{N}$ is a scattering amplitude that can be obtained by fusing amplitudes ${S}_{I}$ and ${S}_{J}$ due to the bootstrap prescriptions outlined in sections 2.3 and 4.2, and $\zeta = {\alpha}_{n}\slash2$  is the constant related to the pole structure of the amplitude (\ref{bootdiag}).   
This equation is valid, in particular, for the amplitudes involving bound states in the sine-Gordon model that we considered earlier. Now we translate it to the space of celestial amplitudes, assuming that $\mathcal{A}_N\left(\omega\right)$ corresponds to ${S}_{N}$, $\;\mathcal{A}_I\left(\omega\right)$ corresponds to ${S}_{I}$ and $\mathcal{A}_J\left(\omega\right)$ corresponds to ${S}_{J}$ via the general definition (\ref{fourier_ampl}).

Transforming (\ref{bootdiag_1}) and converting it to convolution, we get
\begin{multline}
	\mathcal{A}_N\left(\omega\right)=\int_{-\infty}^{\infty} e^{i\omega\theta} S_N\left(\theta\right)d\theta\\
	\shoveleft{=\int_{-\infty}^{\infty} e^{i\omega\theta} {{S}_{I}}\left( \theta -\zeta \right){{S}_{J}}\left( \theta +\zeta \right) d\theta}\\
	\shoveleft{=\int_{-\infty}^{\infty} e^{i\omega\left(\theta^\prime+\zeta\right)} {{S}_{I}}\left( \theta^\prime \right){{S}_{J}}\left( \theta^\prime +2\zeta \right) d\theta^\prime}\\
	\shoveleft{=e^{i\omega\zeta}\int_{-\infty}^{\infty} e^{i\omega\theta^\prime} {{S}_{I}}\left( \theta^\prime \right){{S}_{J}}\left( \theta^\prime +2\zeta \right) d\theta^\prime}\\
	=\frac{e^{i\omega\zeta}}{2\pi}\int_{-\infty}^{\infty} e^{i\omega\theta^\prime} {{S}_{I}}\left( \theta^\prime \right) \left[\int_{-\infty}^{\infty}e^{-i\omega^\prime\theta^\prime-i2\zeta\omega^\prime}\mathcal{A}_J\left(\omega^\prime\right) d\omega^\prime\right] d\theta^\prime\\
	=\frac{e^{i\omega\zeta}}{2\pi}\int_{-\infty}^{\infty}e^{-i2\zeta\omega^\prime} \mathcal{A}_J\left(\omega^\prime\right) \mathcal{A}_I\left(\omega-\omega^\prime\right)d\omega^\prime.
\end{multline}

Therefore, we obtain an additional relation in the space of celestial amplitudes corresponding to the fusion rule for the integrable S-matrices. The aim of translating analytical constraints such as crossing, unitarity and fusion bootstrap rule into celestial space is to bootstrap celestial amplitudes, that is, determine the correlators purely out of analytical conditions. 
The full self-consistent framework of analytical constraints sufficient to determine the higher order celestial amplitudes from the lower order ones has not yet been established.

\section{Conclusion}

The bootstrap method is widely used for various integrable theories, including supersymmetric theories and integrable models in higher dimensions. Modern applications of bootstrap techniques have arised in such theories as $N=4$ supersymmetric Yang-Mills \cite{n4_sym}, relativistic $AdS_2$ \cite{bads}, three-dimensional Ising model \cite{bising}, and others \cite{b2dim}, as well as deformations of integrable two-dimensional theories \cite{deform}. The bootstrap method is a perspective approach to the exact computation of scattering amplitudes, form factors and conformal correlation functions.

In this work, we have provided a detailed description of the S-matrix bootstrap technique in application to the sine-Gordon model, starting from the axiomatic algebraic formalism. We have calculated scattering amplitudes involving higher bound states by fusing the amplitudes for lower bound states and fundamental ones and obtained original integral expressions for the sine-Gordon bound state scattering amplitudes. We have performed consistency checks and demonstrated the equivalence of different representations of the amplitudes.

We have also considered applications of integrable S-matrices in the context of calculating correlation functions. We have demonstrated the direct connection between certain integrable representations of scattering amplitudes and form factors in two-dimensional integrable theories and computed minimal form factors for the sine-Gordon model from our results for integral scattering amplitudes. We suggest a deeper investigation of this connection, including its modification in accordance with other possible representations of amplitudes. Another important field of further research is the relation between scattering amplitudes and generalized form factors on the structural and algebraic level, involving operator identities, complex pole structure and advanced techinques such as Bethe ansatz  \cite{ffsinh,sff}.

We have addressed applications of the two-dimensional integrable $S$-matrices to the calculation of conformal correlators known as celestial amplitudes, which is a rapidly developing area of current investigations in the context of quantum qravity. Particular simplifications arise in this correspondence for the two-dimensional integrable case, where the celestial amplitude becomes a Fourier transform of the rapidity-parametrized S-matrix. We have provided the formulation of the crossing and unitarity conditions specific for the sine-Gordon model, as well as the bootstrap fusion equation, in the celestial Fourier space. Therefore, we have obtained further analytical constraints for celestial amplitudes corresponding to two-dimensional models. Providing a set of constraints sufficient to obtain the higher order conformal correlators from the lower order ones remains an open question. Non-perturbative bootstrap for celestial amplitudes could lead to a wider investigation of conformal symmetries in quantum gravity and remains a promising field of study. We also suggest the application of this technique to other known integrable S-matrices possessing $O\left(N\right)$ symmetry.

\bigskip

\section{Acknowledgenements}

The author is indebted to I. E. Shenderovich for formulating the problem and supervision at the initial stage of the project, and to A. G. Pronko for reading the manuscript and critical remarks. This work has been supported in part by the grant 075-15-2022-289 contributed to the Euler International Mathematical Institute.

\end{document}